\documentclass[11pt,twoside]{article}
\title{Quantum gravity models - a brief conceptual summary\thanks{Invited article for the book ``Mathematical Structure of the Universe'', publ. Copernicus Center Press, Copernicus Center for Interdisciplinary Studies, Cracow 2014, pp. 277-300.}}

\author{Jerzy Lukierski}
\date{University of Wroc\l aw}

\usepackage{amssymb}
\usepackage[usenames,dvipsnames]{color}
\usepackage{amsxtra}
\usepackage{subeqnarray}
\usepackage{sectsty}
\usepackage{titling}
\usepackage{patchcmd}
\usepackage{geometry}
\usepackage{lettrine}
\usepackage{setspace}
\usepackage{harvard}
\usepackage[T1]{fontenc}
\harvardparenthesis{square}
\usepackage{amsmath,amssymb,amscd,amsthm}

\catcode`@=11
\def\secteqno{\@addtoreset{equation}{section}
\def\theequation{\thesection.\arabic{equation}}}
\catcode`@=12
\secteqno

\begin{document}



\maketitle
\begin{abstract}
After short historical overview  we describe the difficulties with application of standard QFT methods in quantum gravity (QG).
The incompatibility of QG with the use of classical continuous space-time required conceptually new approach. We present briefly three proposals: loop quantum gravity (LQG), the field-theoretic framework on noncommutative space-time and QG models formulated on discretized (triangularized) space-time. We evaluate these models as realizing expected important properties of QG: background independence, consistent quantum diffeomorphisms, noncommutative or discrete structure of space-time at very short distances, finite/renormalizable QG corrections. We only briefly outline an important issue of embedding 	QG into larger geometric and dynamical frameworks (e.g. supergravity, (super)strings, p-branes and M-theory), with the aim to achieve full unification of all fundamental interactions.
\end{abstract}
\newpage
\subsection*{Contents:}
\begin{description}
\item{1.} Introduction
\item{2.} Problems with Einstein quantum gravity (QG) as quantum field-theoretic model and dynamical noncommutative quantum space-time
\item{3.} Basic QG models
\begin{description}
\item{3.1} General remarks
\item{3.2} Loop QG
\item{3.3} QG as field theory on noncommutative space-time
\item{3.4} Functional formulation of QG and lattice theories
\end{description}
\item{4.} Final remarks
\end{description}



\section{Introduction}
\lettrine[lines=2]{T}{he} first general relativity (gravity) model was formulated as nonlinear field theory of gravitational metric field $g_{\mu\nu}(x)$ and   provided  in 1915 the Einstein field equations  following from the Einstein-Hilbert action

\begin{equation}\label{lukeq1.1}
S^{EH} = - \frac{1}{2\kappa} \int d^4 x \sqrt{{\det g_{\mu\nu}}}\, R \, ,
\end{equation}
where $R=R_{\mu\nu}^{\quad \mu\nu}$ is the scalar curvature  (see e.g. \cite{MTW,lukbibl2}).
The gravitational field is introduced as characterizing the (pseudo)Riemannian geometry of  curved space-time with the invariant length element $ds$ expressed in terms of the local metric $g_{\mu\nu}(x)$ as follows
\begin{equation*}
ds^2 = g_{\mu\nu}(x) dx^\mu \, dx^\nu \, .
\end{equation*}
The action (\ref{lukeq1.1}) is invariant under the local space-time transformations (diffeomorphisms)
\begin{equation}\label{lukeq1.3}
x'_\mu \equiv x'_\mu (x) = x_\mu + \chi_\mu(x) \, .
\end{equation}
The Einstein equations which describe the gravitational field interacting with matter take the following form
\begin{equation}\label{lukeq1.4}
G_{\mu\nu}(x) \equiv R_{\mu\nu}(x) - \frac{1}{2} g_{\mu\nu}(x)\, R(x) = \kappa T_{\mu\nu}(x) \,,
\end{equation}
where $R_{\mu\nu}$ is the Ricci tensor (see e.g. \cite{MTW,lukbibl2}), $\kappa=\frac{8\pi G}{c^2}$  ($G$ -- Newton constant, $c$ -- light velocity) and $T_{\mu\nu}(x)$  the local energy-momentum tensor which describes the  matter distribution in space-time. From (\ref{lukeq1.4}) follows that $T_{\mu\nu}(x)$ determines the  space-time curvature tensor $R_{\mu\nu}^{\quad \rho\tau} (x)$, i.e. gravity is a dynamical theory of curved space-time providing geometric interpretation of matter distribution.

The form (\ref{lukeq1.4}) of Einstein equations leads to the question of how to define its rhs, i.e. how to describe in field-theoretic framework the tensor $T_{\mu\nu}(x)$ representing gravitational  matter sources. The development of field theory provided us however with  the answer of how the matter is described by various fields. In particular,
 matter fields
 were described by
\begin{description}
\item{-- in 1926} -- scalar Klein-Gordon field $\phi(x)$ \cite{lukbibl3}, describing e.g. Higgs particle;
\item{-- in 1926} -- spinorial Dirac field \cite{lukbibl4} $\psi_A(x)$, describing e.g. electron/positon and quarks;
\item{-- in 1954} -- vectorial Yang-Mills (YM) field $A^{i}_\mu$ \cite{lukbibl5}, where the  indices $i$ describe the adjoint representation of an internal symmetry group. By YM fields are described e.g. in QCD, gluons  with internal symmetry $SU(3)$ or vectorial $W$-bosons in electroweak sector of Standard Model (SM) with internal symmetries $SU(2)\times U(1)$ (see e.g. \cite{lukbibl6}).\footnote{It should be mentioned that vectorial EM field $A_\mu(x)$ with $U(1)$ internal symmetry has been known since 1861, when the Maxwell equations describing electrodynamics were introduced.}
\end{description}

Relativistic  free fields describing noninteracting elementary particles are cha\-racterized    by mass $m~\geqslant~0$ and spin $s$ ($s=0,\frac{1}{2},1, \ldots$) \cite{lukbibl7}, with the listed above three examples of fields corresponding to $s\leqslant 1$.
 We add that the gravitational field $g_{\mu\nu}(x)$ corresponds to the choice $m=0$ and $s=2$.

The first effort to unify all fundamental interactions was successfully undertaken inside the local $D=4$ field-theoretic framework in the 1970's by coupling the gravitational sector to the collection of fields describing  elementary particles in the framework of the Standard Model (SM).
The action describing all elementary processes in Nature was proposed in the following form
\begin{equation}\label{lukeq1.5}
S^{\rm tot}= S^{\rm EH} + S^{\rm SM} + S^{\rm int} \, ,
\end{equation}
where $S^{\rm SM}$ is the field-theoretic action of the SM and $S^{\rm int}$ describe its  gravitational interactions, with the rhs of (\ref{lukeq1.4}) defined by the variation of $S^{\rm int}$ under the change $g^{\mu\nu} (x) \ \to \ g^{\mu\nu} (x) + \delta g^{\mu\nu} (x)$, i.e.
\begin{equation*}
T_{\mu\nu}(x) = \frac{\delta S^{\rm int}}{\delta g^{\mu\nu}(x)}\,.
\end{equation*}

The next step  is to introduce the quantization of all  fields which occur in the action (\ref{lukeq1.5}). It was realized  that in order to calculate the scattering processes of elementary particles one should use the perturbative methods and consider perturbative expansions in  quantum  free fields which introduce the algebra of field oscillators.
After introducing the Higgs mechanism \cite{lukbibl8}, which generated the
 mass for the appropriate gauge fields,  it  appeared that in the quantized SM model, the perturbative calculations performed via Feynman diagrams techniques lead to  the formulae for scattering amplitudes with infinities
  which one cannot
 remove by renormalization procedure.
Unfortunately, the Einstein gravity, described by the action (\ref{lukeq1.1}), after quantization leads in perturbative calculations to infinities which are  not  renormalizable.

\section{Problems with Einstein quantum gravity (QG) as a quantum field-theoretic model and dynamical noncommutative quantum space-time}

From a logical point of view it is still not excluded that gravity in Nature remains classical.
Even if in the action (\ref{lukeq1.5}) we quantize all the fields of SM, the requirement  that the classical gravitational background has to be promoted to dynamical quantum field cannot be formally proven (see e.g. \cite{lukbibl9}). This option is however usually not accepted \cite{lukbibl10}. It  also appeared that quantized Einstein gravity can be introduced as a suitably constrained  gauge QFT of massless spin two field with local diffeomorphism (\ref{lukeq1.3}) as the local gauge group (see e.g. \cite{lukbibl11}).
  Then, it was advocated that the aim of the theory of fundamental interactions is to incorporate the gravitational degrees of freedom in the field-theoretic quantum `Theory of Everything', containing all quantized fields.

Two further ways of dealing with the   unsurmountable property of  nonrenormalizability of perturbative Einstein QG were assumed in the standard QFT approach:

\begin{description}
\item{\textbf{-- }}
  To apply the methods of renormalization group and introduce a running energy-dependent gravitational coupling constant scenario with nonzero fixed point. Such so-called `asymptotic safety' approach was initiated  by Weinberg in the 1970's \cite{lukbibl12} and further developed with some success by other authors  (see e.g. \cite{lukbibl13,lukbibl14}).

 \item{\textbf{-- }} To modify the Einstein-Hilbert action, in particular by the following replacement in  (\ref{lukeq1.1})
 \begin{equation}\label{lukeq2.1}
 R \ \to \ R + aR^2 + b R_{\mu\nu}R^{\mu\nu} +cR_{\mu\nu\rho\tau}R^{\mu\nu\rho\tau}+ \ldots \,.
 \end{equation}
 \end{description}
Various particular cases of formula (\ref{lukeq2.1}) were studied, for example:
\begin{description}
\item{$\bullet$} $c=0$ -- Stelle family of actions \cite{lukbibl15}, providing renormalizable theory but after quantization containing the ghost states\footnote{The ghost states are described by quantum states with zero or negative norm, something which is not consistent with the basic postulates of QM.}.
\item{$\bullet$} $b=c=0$ -- Starobinsky model \cite{lukbibl16} which was successfully applied in the construction  of the cosmological model describing inflation as well as nowadays  the  acceleration of the expansion of the Universe.
\item{$\bullet$} $ a=c =  \frac{1}{4} b$ - Zwiebach `stringy' gravity model inspired by quantum corrections in string theory \cite{lukbibl17}.
\end{description}

In the class of modified Einstein QG models  within the standard QFT approach one should include the recent modifications of Einstein theory with an asymmetric treatment of space and time, which lead to  renormalizable class of Ho\v{r}ava-Lifschitz gravities \cite{lukbibl18}, as well as to so-called $f(R)$ gravities (see e.g. \cite{lukbibl19,lukbibl20}).

Further in this presentation we shall be  interested in QG models formulated  outside the standard QFT approach, with new proposals for the  construction of the quantum counterparts of standard classical gravity framework.
For such a purpose it is important to single out the basic reasons for the failure of standard QFT approach to QG.

 There are two important difficulties with applying the standard  quantum field-theoretic methods to QG:
 \\
 \indent{\textbf{i)} \textsl{Technical problem:} The nonlinearity of Einstein-Hilbert action and the dimensionfull nature of gravitational coupling constant $\kappa \sim G$ leads to $D=4$ perturbative nonrenormalizability. It appears that  the action densities describing  renormalizable $D=4$ local field theories are low-order (up to fourth) polynomials in local  field variables.
 \\
 \indent{\textbf{ii)} \textsl{Conceptual problem:} The feature that QG describes the quantized geometrodynamics of space-time is alien to the framework of standard classical and quantum field theory. In the quantization procedure the classical space-time
  with numerical coordinates $x_\mu$ due to QG effects becomes dynamical,  noncommutative
 \begin{equation*}
 \begin{array}{c}
 \hbox{classical}
 \\
 \hbox{space-time} \ x_\mu
 \end{array}
 \qquad
 {{\xrightarrow[{\phantom{ssssss}}]{QG}}}
 \qquad
 \begin{array}{c}
 \hbox{quantum}
 \\
 \hbox{space-time}\ {\hat{x}}_\mu
 \end{array}.
 \end{equation*}
 We recall that the standard quantum field is  described on flat (static) Minkowski space $x_\mu \in {\mathcal M}^{3,1}$ as the Fourier momentum  integral over the   quantized field  oscillators $\hat{a}(p)$
 \begin{equation*}
 \widehat{\phi} (x)\, = \,  \int d^4 p \,{\mu (p)}\, \hat{a} (p) e^{ipx} \, ,
 \end{equation*}
 where
   the measure $\mu(p)$ for free fields incorporates the mass-shell condition. In the presence of QG effects, the classical space-time  describing field arguments is replaced by the quantum space-time, i.e. the standard quantum fields
    due to QG are modified as follows\footnote{The quantum space-time is usually characterized by noncommutative or discretized commutative space-time coordinates.}
\begin{equation*}
\widehat{\phi}(x) \quad
{{\xrightarrow[{\phantom{ssssss}}]{QG}}}
\quad
\widehat{\phi}(\hat{x}) \,.
\end{equation*}

It has been shown \cite{lukbibl22,lukbibl23,lukbibl24,lukbibl21} that in the presence of a quantized gravitational field one cannot measure two different space-time coordinates with arbitrary accuracy.
 Assuming that:
\\
{\textbf{i)}} gravitational field satisfies the Einstein equations (see \ref{lukeq1.4}) and
\\
{\textbf{ii)}} quantum-mechanical Heisenberg uncertainty relations
 restricting the measurements of position/three-momenta and energy/time
\begin{subeqnarray}\label{lukeq2.5}
&&\Delta x_i \, \Delta p_j \geqslant \delta_{ij}\,\hbar \, , \\
&& \Delta t \, \Delta E \geqslant \hbar
\end{subeqnarray}
are valid, one can derive that 
\begin{equation}\label{lukeq2.6}
\Delta x_i \, \Delta x_j \geqslant \delta_{ij}\, l^2_p \, ,
\qquad \qquad
l_p = \sqrt{\frac{\hbar G}{c^2}}\,  \simeq 10^{-33}\hbox{cm} \, .
\end{equation}
Further, due to (\ref{lukeq2.5}) and relativistic covariance,  the relations (\ref{lukeq2.6}) can be extended to the following relativistic form of the uncertainty relations
\begin{equation}\label{lukeq2.7}
\Delta x_\mu \, \Delta x_\nu \geqslant l^2_p \qquad \qquad (\mu\neq \nu) \, .
\end{equation}
In the algebraic formulation that leads to (\ref{lukeq2.7}) one can equivalently
 postulate the following noncommutativity of quantum space-time coordinates
\begin{equation}\label{lukeq2.8}
\left[ \hat{x}_\mu , \hat{x}_\nu \right] = l^2_p\, \theta^{(0)}_{\mu\nu} +
\ldots = l^2_p\, \theta_{\mu\nu} \left( \frac{\hat{x}}{l_p}\right) \,,
\end{equation}
where $\theta^{(0)}_{\mu\nu}= -\, \theta^{(0)}_{\nu\mu}$ is a constant antisymmetric tensor.
In the general case the antisymmetric operator-valued function $\theta_{\mu\nu}(\hat{x})$ can be generalized to  a function of the quantum phase space coordinates $(\hat{x}_\mu, \hat{p}_\mu)$, namely\footnote{In such a case we should supplement (\ref{lukeq2.8}) with the formulae for the commutators $[ \hat{x}_\mu, \hat{p}_\nu ]$ and $[ \hat{p}_\mu, \hat{p}_\nu ]$ in a way consistent with Jacobi identities.}
\begin{equation}\label{lukeq2.9}
\theta_{\mu\nu}(\hat{x}) \ \rightarrow \ \Theta_{\mu\nu}(\hat{x},\hat{p}) \, .
\end{equation}
Various models of quantum space-times and quantum phase spaces
  are characterized by different choices of $\theta_{\mu\nu} (\hat{x})$ or
 $\Theta_{\mu\nu}(\hat{x},\hat{p})$.
 It should be added that the relations (\ref{lukeq2.8}) for any  $\theta_{\mu\nu} (\hat{x})$ necessarily break the standard Lorentz invariance, however, as it was shown firstly by Snyder
  \citeyear{lukbibl25},   if we choose particular momentum-dependent noncommutativity
  \begin{equation*}
  \Theta_{\mu\nu}(\hat{x}, \hat{p}) = M_{\mu\nu} = \hat{x}_\mu \hat{p}_\nu - \hat{x}_\nu \hat{p}_\mu \, ,
  \end{equation*}
  we get the Lorentz-invariant Snyder model of quantum space-time.

  From the mathematical point of view, quantum space-time is a noncommutative algebraic manifold, with a basic relation (\ref{lukeq2.8}) (see also (\ref{lukeq2.9}) and footnote${}^{3}$). Physically, the appearance of a noncommutative
  structure can be explained by the gravitational effect of creating
   microscopic black holes due to the appearance of very high energy densities in the region where the  space-time localization measurement is performed (such an effect was predicted firstly by M.P.~Bronstein \citeyear{lukbibl21}).
  In consequence the QG effects lead to the atomization of physical space-time at very short distances  .
  $$
  \begin{array}{c}
  \hbox{reaction of dynamical}
  \\
   \hbox{space-time to quantum}
   \\
    \hbox{measurement process}
  \end{array} \quad \longleftrightarrow \quad
  \begin{array}{c}
   \hbox{below $10^{-33}$cm the notion}
   \\
    \hbox{of classical space-time}
    \\
     \hbox{looses operational meaning}
  \end{array}
  $$

In the physical models formulated in noncommutative space-time, the uncertainty relations (\ref{lukeq2.7}) are incorporated on a basic algebraic level i.e. the limitations of local space-time measurements are becoming an `ontological' property of such an approach.

In the framework with quantum space-time, the Planck length or Newton constant becomes, besides $c$ and $\hbar$, the third fundamental geometric parameter.
 In particular, one can introduce the following three frameworks describing three basic domains of physical dynamical theories:

\medskip
\begin{center}
\fcolorbox{black}{white}{
\begin{minipage}[t]{10cm}
\begin{center}Classical mechanics and classical field
theory -- \newline $\hbar=0$, \ $G=0$;  ($x_\mu$, $p_\mu$) commutative
\end{center}
\end{minipage}
}

$$ \Downarrow \hbar \neq 0$$

\fcolorbox{black}{white}{
\begin{minipage}[t]{10cm}
\begin{center}Quantum mechanics and standard QFT --
\newline $\hbar \neq 0, \ G=0$;  standard quantum  phase space
\newline ($\hat{x}_{\mu}, \hat{p}_\mu$) with commuting space-time \end{center}
\end{minipage}
}

$$ \Downarrow G \neq 0$$

\fcolorbox{black}{white}{
\begin{minipage}[t]{10cm}
\begin{center}QM and  QFT  in presence of QG effects --
\newline $\hbar \neq 0, \ G\neq0$;
deformed  quantum  phase space \newline
 ($\hat{x}_{\mu}, \hat{p}_\mu$)
 with noncommutative space-time \end{center}
\end{minipage}
}
\end{center}

The third sector ($\hbar\neq 0, G\neq 0$) of quantum theories containing the QG effects  can be described by new mathematical tools based on the notions of algebraic geometry:
 quantum groups, noncommutative geometries and quantum spaces (see e.g. \cite{lukbibl26,lukbibl27,lukbibl28,lukbibl29}).

Similarly as the existence of energy quanta follows from basic postulates of quantum mechanics, the quantum space-time geometry implies that at very short distances we are dealing with quanta of space-time heuristically illustrated in $D=4$ by cells in space-time with volume $\sim l^4_p$.
This picture leads to a popular conclusion that due to QG effects the space-time at very short Planck lengths has a discrete structure.

\section{Basic QG models}
\subsection{General remarks}

The modification at short distances of classical space-time structure distinguishes new QG models from more conventional treatments of QG which propose embedding of gravity into a larger geometric structure, with the subsector of standard classical space-time  and standard canonical  phase space coordinates.

In the conventional approaches to QG the construction of a finite (renormalizable) unified theory is achieved by introducing additional degrees of freedom coupled to gravity sector in order to `smooth' the infinities occurring as nonrenormalizable results of  quantum calculations.
The first efforts towards such extended space-time framework were based on a Kaluza-Klein scenario, with space-time enlarged by new $n$ dimensions ($D=4 \to D=4+n$; usually  $1\leqslant n \leqslant 7$) (see e.g. \cite{lukbibl30}).
Due to the appearance of bosons and fermions in the observable spectrum of elementary particles, the geometric programme of  the unification of elementary interactions  naturally leads to  supersymmetric theories, with $D=11$ supergravity as the first Theory of Everything  unifying  all  interactions \cite{lukbibl31}.
The description which requires $D>4$ as well as supersymmetry was further conceptually  widened  by the introduction of extended elementary objects
 with mostly studied (super)string theory.
    Chronologically, the next Theory of Everything containing QG was described by $D=10$ superstrings \cite{lukbibl32} providing after second quantization (i.e. in the framework of quantum string theory) the
   finite renormalization corrections in its gravity sector.

At present, QG is studied in the literature in basically two ways:

-- As embedded in covariant theory of extended $p$-dimensional elementary objects, called $p$-branes ($p=1,2, \ldots, D-1$). In such approach the (super)string  formalism  is generalized to so-called $M$-theory formulated in $D=11$ \cite{lukbibl33}, which is considered also as the actual Theory of Everything.

-- By introducing nonstandard new models of quantized Einstein gravity. \linebreak These models are supposed to provide finite (or renormalizable) QG by modifying the canonical field-theoretic approach
 and lead to the various
ways of dealing with  quantum space-time.

In this article we shall restrict our comments to the following three basic nonstandard QG approaches:
\begin{description}
\item{$\bullet$} Loop quantum gravity (LQG) (see e.g. \cite{lukbibl34});
\item{$\bullet$} QG formulated as QFT on noncommutative space, which may be equivalently described as a particular class of nonlocal field theories on standard space-time (see e.g. \cite{lukbibl36,lukbibl37});
\item{$\bullet$} Lattice models of QG which are linked  with the functional formulation of QG  by the use of discretized space-time and its triangulations  (see e.g. \cite{lukbibl38}).
\end{description}

We shall discuss the QG models by taking into  consideration  the following desirable features:
\begin{description}
\item{A.} The description should not be attached  to any particular choice of space-time, i.e. the formulation should be background-independent.

\item{B.} The model should incorporate a quantum counterpart of the diffeomorphism invariance.

\item{C.} At very small distances
   the quantum space-time structure should appear to be
    described by  noncommutative or commutative but discrete manifold.

\item{D.} The model should provide renormalizable, or even better  finite,    formulae   which describe the quantum fundamental processes involving QG.
\end{description}

Various new QG models address these requirements in different ways. We mention that we did not include here as a fifth desired property, namely the requirement of the unification of all fundamental interactions, however it is quite plausible that
 the formulation of a finite unified theory will be related with the embedding of QG in a  new, as yet  unknown,   quantum Theory of Everything.

\subsection{Loop quantum gravity (LQG)}

Einstein gravity based on Einstein-Hilbert action (\ref{lukeq1.1}) is a nonlinear local field theory of a fundamental gravitational field, with  local diffeomorphism invariance. As early as the 1960's  (see e.g. \cite{lukbibl39}), by exploiting the analogy between local diffeomorphisms and local non-Abelian internal symmetries in
Yang-Mills theories  a new kinematic description of gravity
 was conjectured
  by introducing the nonlocal loop (or holonomy) variables  ($\mathcal L$ - closed loop; $i=1,2,3$)
\begin{equation}\label{lukeq3.1}
A^i [ {\mathcal L};t ] = \exp \oint\limits_{\mathcal L} A^i_r (\vec{x}, t)dx^r \qquad \quad (r=1,2,3)\,,
\end{equation}
which are invariant under local three-dimensional space diffeomorphism transformations.
The generalized field coordinates (\ref{lukeq3.1}) were further supplemented by generalized field momenta described by nonlocal flux variables ($S$ -  closed surface; $i,j,k =1,2,3$)
\begin{equation}\label{lukeq3.2}
E^i [ S; t] = \varepsilon^{rst} \oint\limits_{S} E^i_r (\vec{x}, t) dx_s \wedge dx_t \, .
\end{equation}
The fields $A^i_r(x)$, $E^i_r(x)$ are the conjugate Ashtekar SU(2) gauge variables \linebreak  \cite{lukbibl39a}
 which express  locally the curved space metric  tensor field $g_{rs} (x)$ ($r,s=1,2,3$)
 by means of the
  triad $E^i_r$ (`Ashtekar electric field') as follows
\begin{equation*}
g_{rs} = \sum\limits_i E^i_r \, E^i_s \cdot (\det E)^{-1}\, .
\end{equation*}
The variables (\ref{lukeq3.1}--\ref{lukeq3.2}) define new phase space with holonomy-flux Poisson algebra which provides the quantization rules

\begin{equation}\label{lukeq3.4}
\left\{
A^i [ {\mathcal L}; t ] , E^i [S; t ]
\right\}
\ {{\xrightarrow[{\phantom{ssssss}}]{\hbar \neq 0}}}  \
\left[
\hat{A}^i [ {\mathcal L}; t ], \hat{E}^i [S; t]
\right] \,.
\end{equation}

The description of LQG  by basic variables (\ref{lukeq3.1}--\ref{lukeq3.2}) is  linked with  the canonical approach -- i.e. the nonlocal phase-space coordinates are defined nonlocally in space but  for fixed `local' value of global time $t$.
  The quantization rules (\ref{lukeq3.4}) are in principle  derivable from the canonical formulation of Einstein gravity. In such new  phase-space  formulation  of  gravity the dynamics is provided by 
  the following kinematic and dynamical constraints \cite{lukbibl34}
\begin{description}
\item{$\rightarrow$} three SU(2) (Gauss) kinematic constraints,

\item{$\rightarrow$} three kinematic constraints generated by space diffeomorphisms,

\item{$\rightarrow$} dynamical Hamiltonian constraint following from the covariance under the time translations (time reparametrisations).
\end{description}

In LQG, the local field operators describing standard formulation of Einstein gravity are replaced  by the functionals depending on the loop variable (\ref{lukeq3.1})
\begin{equation*}
g_{\mu\nu} \quad \longrightarrow \quad \Phi \left[ A^i [ {\mathcal L} ]; t\right] \, .
\end{equation*}

The standard quantization
  with excitations described by the field-theoretical creation/anihilation operators
 (e.g. gravitons realized in Fock space) is replaced in LQG by the quantized functionals $\Phi$
acting on
  nonseparable `polymer' Hilbert spaces which are  spanned by the superposition of SU(2) spin network states.
These states form a quantum spin lattice and in particular permit to calculate (see e.g. \cite{lukbibl34})
\begin{description}
\item{$\bullet$} the spectrum of areas describing elementary quantum surface elements, proportional to $\gamma l^2_p$ ($\gamma$\ -\ Barbero-Immirzi parameter);
\item{$\bullet$} the spectrum of elementary quantum volume elements ($\sim l^3_p$).
\end{description}

Both area and volume spectra are bounded from below and introduce dynamically determined granular structure of space, with minimal surfaces and minimal volumes defining the sizes  of `space quanta'.

Subsequently it was proposed a spacetime covariant extension of canonical LQG -- the covariant LQG framework, with nonrelativistic SU(2) spin network states replaced by space-time lattices and SU(2) algebra indices replaced by SL(2;C) labeling.
The covariant LQG approach is closely linked with so-called spin foam models (see e.g. \cite{lukbibl40}), with the space of states defined by the spin foam graphs. It can also be argued that the covariant LQG model is closely linked to the four-dimensional lattice approach to QG which will be considered in Sect.~3.3.

One can summarize the properties of the LQG approach by providing the following answers to the list of desirable properties of QG models (see A-D in Sect.~3.1).
\begin{description}
\item{{\textbf{A.}}} The LQG formulation is background-independent by construction; the notion of quantum space-time emerges in a dynamical way.

\item{{\textbf{B.}}} The diffeomorphism-invariant spaces are obtained by introducing the equivalence classes of graphs defining the spin network states \cite{lukbibl34}.

\item{{\textbf{C.}}} An advantage of LQG approach is the dynamical  derivation of discrete  noncommutative structure of quantum space-time,  described in particular by minimal surfaces and minimal volumes.

\item{{\textbf{D.}}} The LQG calculations,  due to the appearance of dynamical discrete structure which is the physical regularization,  should lead to finite results, with lattice lengths determined  by gravity dynamics. At present, however, the explicit LQG calculations provide  rather modest results.
\end{description}

One can add  that there are still problems with the description in LQG framework of the classical  general relativity results, in particular the derivation of  known forms of Einstein equations and gravitational classical  solutions.
Also, in LQG formalism  the description of gravitons as quantum   excitations requires nontrivial dynamical considerations.
On the other hand, the  simplified one-dimen\-sional  FRW models have been successfully quantized and applied in cosmology.
 Using such `toy' cosmological  LQG models it has been deduced that the evolution of the Universe need not begin with a primary singularity, i.e. the Big Bang  postulate appears not to be supported by the LQG approach.

\subsection{QG as QFT on noncommutative space-time}

In such an approach one inserts in QG geometric framework explicitly the noncommutativity of space-time. In place of standard commuting space-time coordinates $x_\mu$ one uses the following noncommutativity  relations  for generators $\hat{x}_\mu$ of quantum space-time  (see also (\ref{lukeq2.8}))

\begin{equation}\label{lukeq3.6}
[ x_\mu , x_\nu ]
=0 \quad \Rightarrow \quad
[ \hat{x}_\mu , \hat{x}_\nu ] = i l^2_p \, \theta_{\mu\nu}
\left( \frac{\hat{x}}{l_p} \right) \, ,
\end{equation}
where the operator-valued tensorial field $\theta_{\mu\nu}$ can be expanded in the gravitational coupling constant as the power expansion $\displaystyle \bigg( y= \frac{\begin{Large}\hat{x} \end{Large}}{l_p} \bigg)$
\begin{equation}\label{lukeq3.7}
\theta_{\mu\nu}(y) = \theta_{\mu\nu} + \theta_{\mu\nu}{}^{(1) \, \rho} y_\rho + \theta_{\mu\nu}{}^{(2)}{}^{\rho\tau} y_\rho y_\tau + \ldots \, ,
\end{equation}
where $\theta_{\mu\nu}{}^{(n) \rho_1 \ldots \rho_n}$ are constant tensors.

The first constant term defines canonical (called also (DFR) Dopplicher\-Fredenhagen-Roberts) deformation \cite{lukbibl23};
the second one describes Lie-algebraic deformation of space-time algebra, which in particular includes quite popular   $\kappa$-deformation \cite{lukbibl41,lukbibl42,lukbibl43}.
The relations (\ref{lukeq3.6}) in quantum phase space formulation can be generalized by assuming that the commutator
$[ \hat{x}_\mu, \hat{x}_\nu ]$ depends also on momenta or even other dynamical primary variables (see e.g. \cite{lukbibl44}).

It should be added that at the turn of the millennium there \cite{lukbibl44a,lukbibl44b} the modification of special relativity called `Doubly Special Relativity' (DSR) with Planck mass as (in addition to light velocity $c$) the second kinematic invariant was introduced.
The postulates of DSR, linked with the earlier description of quantum $\kappa$-deformed Poincar\'{e} symmetries \cite{lukbibl44c,lukbibl44d}, became in recent years a fertile incentive for proposing the calculations aiming at the description of possible QG effects (see e.g. \cite{lukbibl44e,lukbibl44f}).

In the approach using noncommutative (quantum) space-time the QG corrections are introduced into field theory in an algebraic way,  by the substitution of the
   classical space-time coordinates  $x_\mu$
   in standard quantum fields (e.g. describing Standard Model) by the noncommutative   $\hat{x}_\mu$.
In such a way one can introduce the noncommutative classical fields
 $\phi_A (\hat{x})$ and the noncommutative quantum fields
 $\hat{\phi}_A (\hat{x})$.
  If we take into consideration the  QG corrections, the quantization of classical fields is obtained by the substitution
 \begin{equation*}
 \phi_A ({x}) \sim \sum\limits_p a(\vec{p}) \,e^{ipx} \quad
 \rightarrow \quad
 \hat{\phi}_A (\hat{x})
 \sim \sum\limits_p \hat{a}(\vec{p}) \,e^{ip\hat{x}} \, ,
 \end{equation*}
 with the noncommutativity of quantum fields $ \hat{\phi}_A (\hat{x})$  introduced
  by quantized field oscillators as well as
  by the  noncommutative   space-time coordinates.

 The local  multiplication of  classical and quantum fields
   modified by QG corrections
  can be realized by the introduction of a nonlocal star multiplication of corresponding standard local fields  (see e.g. \cite{lukbibl44g,lukbibl44h}).
 \begin{equation}\label{lukeq3.9}
 \begin{array}{llll}
 \phi(\hat{x}) \cdot \chi_(\hat{x})
 \quad
 &
{{\xrightarrow[{\phantom{ssssss}\hbox{\footnotesize{map}}}\phantom{ssssss}]{\hbox{\footnotesize{Weyl}}}}}
\quad
 &
 \phi(x) \star \chi(x)
 \qquad
 &
 \hbox{{(classical)}}
 \\[10pt]
 \hat{\phi}(\hat{x}) \cdot \hat{\chi} (\hat{x})
 \quad
 & {{\xrightarrow[{\phantom{ss}\hbox{\footnotesize{Weyl map}}}\phantom{ss}]{\hbox{\footnotesize{(quantum)}}}}}
 \quad
 &
 \hat{\phi}(x) \star \hat{\chi}(x) \qquad
 & \hbox{(quantum)}
 \end{array}
 \end{equation}
The Weyl maps (\ref{lukeq3.9}) provide homomorphisms of noncommutative field algebras, and replace the local noncommutative field theory with some specific nonlocal field theory on standard Minkowski space.
Effectively one can obtain such  noncommutative QG-modified field theory by introducing in classical action the $\star$-product multiplication of fields.
For example, in such an approach the QG corrections in QED will be described
 by the following replacement of standard Maxwell action density
\begin{align*}
F_{\mu\nu} (x) \, F^{\mu\nu}(x) + e J_\mu (x) A^\mu (x)
&&
{{\xrightarrow[\phantom{ss}]{\hbox{\footnotesize{QG corrections}}}}}
&&
F_{\mu\nu} \star F^{\mu\nu} + e J_\mu (x)\star A^\mu (x)
\end{align*}
where $J_\mu (x)$ denotes the external  electric current.

The $\star$-multiplication for simplest canonical (DFR) noncommutative deformation is represented by the Moyal $\star$-product \cite{lukbibl45}.
In such a case it was shown \cite{lukbibl46} that for QED  only the vertices are modified
  in perturbative Feynmann diagrams.

The description of QG corrections in particle physics can be
 described therefore in two equivalent ways

 $$
 \begin{array}{c}
 \hbox{local field-theoretic}
 \\
 \hbox{SM on noncommutative}
 \\
 \hbox{space-time}
 \end{array}
 \qquad
 \leftrightarrow
 \qquad
 \begin{array}{c}
 \hbox{nonlocal SM on}
 \\
 \hbox{classical space-time}
 \\
 \hbox{with $\star$-product of fields}
 \end{array}
 $$

 The noncommutativity (e.g. the tensors $\theta_{\mu\nu}{}^{(n) \rho_1 \ldots \rho_n}$  in (\ref{lukeq3.7})) should be determined dynamically by QG,  e.g.   in standard canonical approach  it should depend on  three-dimensional 2-tensors fields $h_{ij}(\vec{x}, t)$,  $p_{ij}(\vec{x}, t)$ (see e.g. \cite{MTW,lukbibl2}).
 In particular in formula (\ref{lukeq3.6}) the rhs should appear  as some dynamically determined  functional of quantized gravitational canonical fields
 \begin{equation}\label{lukeq3.11}
 \theta_{\mu\nu}(\hat{x}) \qquad \longrightarrow \qquad
 \theta_{\mu\nu} [ \hat{h}_{ij}, \hat{p}_{ij}; \hat{x} ] \, .
 \end{equation}
 Unfortunately the dynamical equations specifying
  the functional  noncommutativity factors (\ref{lukeq3.11}) are not known. As another option   (see e.g. \cite{lukbibl47})
   it has also been postulated that the fields $\theta_{\mu\nu}$ introduce new degrees of freedom in QG.

 From a geometric point of view one can consider
  the nonconcommutative approach
 to QG as described by the formalism of noncommutative Riemannian geometry.
  In such  a framework QG is described by the noncommutative deformation of standard Riemannian geometry, with the use of noncommutative counterparts of basic notions of differential geometry (connections, curvatures, diffeomorphism maps) (see e.g.  \cite{lukbibl48,lukbibl49}).
 It can be added that the noncommutative curved space-times has been also linked with the quantum group approach to  local diffeomorphism generated by $\infty$-dimensional quantum group, with twist  deformation of standard Riemannian geometry described by Drinfeld twist \cite{lukbibl50}.

 There have already been numerous papers with the calculation of noncommutative corrections to Einstein action obtained for some particular classes of deformations (some classes of $\star$-products) (see e.g. \cite{lukbibl51,lukbibl52,lukbibl53}).
 It is interesting to mention that for two basic deformations -- canonical (DSR) and $\kappa$-deformation -- the leading corrections are of second order. i.e.  proportional to $l^2_p$,  far below the present observability threshold.

 The QG approach based on noncommutative space-time geometry provides the following responses to the list of desirable properties listed as A--D in Sect.~3.1.
 \begin{description}
 \item{A.\ \ }
 At present, the noncommutative framework does not provide a background--independent approach, because  the noncommutative space-time structure is postulated.
 The link of noncommutativity with  QG dynamics and possible noncommutative background-independent formulation  is
  according to our opinion (see also \cite{lukbibl53a})
  one of the major challenges in QG.
 It requires the discovery of new fundamental QG equations that determine dynamically the noncommutativity of QG-deformed coordinates and fields.

 \item{B.\ \ } In noncommutative  Riemannian geometry the standard  diffeomorphisms are becoming  quantum operator-valued diffeomorphisms.
 If the deformation of Riemannian geometry is generated by a twist, one can introduce  explicit formulae describing the noncommutative Riemannian geometry (see e.g. \cite{lukbibl50}).
  If one uses $\star$-product formalism, one gets the   calculable  modifications of standard  formalism of general relativity (action, field  equations,  quantum corrections).

 \item{C.\ \ }The noncommutativity of quantum space-time is
  present in this approach as the  consequence of
  a primary postulate.
   This postulate
    should be supplemented with the above-mentioned derivation of noncommutativity from QG dynamics.
 In this respect
  one can mention
    the functional method of integration of gravitational degrees of freedom,  which are
     coupled to  quantum matter, as the way of generating the QG-generated  noncommutativity of matter degrees of freedom (see e.g. \cite{lukbibl54}).

 \item{D.\ \ }In the noncommutative approach to the models of QFT  only in particular cases  some infinities were removed, but a general scheme for the cancellation of  nonrenormalizable divergencies due to the presence of space-time noncommutativity is not known.
   The question   whether  noncommutativity    improves the finiteness of
    the corresponding quantum field
     theory remains  open.
 \end{description}

 \subsection{Functional formulation of QG and lattice approach}

The Feynman quantization method in  canonical field theory leads to the functional integral over the configurations of classical field $\phi$
 \begin{equation}\label{lukeq3.12}
 Z [ \phi ] = \int {\mathcal D} \phi \, e^{i S[ \phi ]} \, ,
 \end{equation}
 where $S[ \phi ]$ describes the action functional determining the field dynamics.
 If $S \equiv S^{\rm EH}$ (see (\ref{lukeq1.1})) one gets QG functional integral \cite{lukbibl55}

 \begin{equation}\label{lukeq3.13}
 Z^{\rm EH} [ g_{\mu\nu} ] = \int {\mathcal D} [ g_{\mu\nu} ] e^{i S^{\rm EH} [ g_{\mu\nu} ]}
 \end{equation}
 describing Einstein QG,
  with functional measure ${\mathcal D} [ g_{\mu\nu} ]$  providing the functional integral over all metrics divided by the diffeomorphisms. One can say alternatively that the formula (\ref{lukeq3.13}) describes the functional over all possible curved space-time geometries represented by diffeo-invariant configurations of the gravitational field.

 It is not known how to calculate explicitly  the functional integral (\ref{lukeq3.12}) unless the action $S [\phi ]$ is free, i.e. linear or quadratic in field variable $\phi$,
   with linear term describing an interaction with external classical current.
 For more general choices of field actions, describing nonlinear interactions, one can however perform the approximate calculations if we discretize the functional integral, i.e. introduce a lattice counterpart of the functional integration.
  In such a way from (\ref{lukeq3.13}) one can derive the lattice formulation of QG.
 Such discretized formulation permits the introduction of new nonperturbative methods for performing QG calculations.
 It should also be added that the lattice discretization is  required if we wish to use advanced computer techniques for the approximate calculations of functional integrals.

 The history of lattice approaches to QG began  more than half a century ago  \cite{lukbibl56}.
  Further substantial progress in lattice methods was   achieved in the framework of QCD calculations \cite{lukbibl57,lukbibl58}, with the  aim of  deriving the properties of hadronic matter from basic theory of its elementary constituents: quarks and gluons.
 The discretization of functional integral in QG is however more complicated than in QCD, because in the theories
 with local internal symmetries  (YM field theory) one can introduce fixed, static  lattices.

 In discretized QG functional integral  however, the lattices  cannot be fixed because after dynamical triangulation procedure  they represent variable geometries which are `latticized'.
  In particular the QG triangulation should provide the lattice version of local coordinate invariance and describe consistently the geometric gravity data, such  as e.g.  space-time curvature.

  There are two versions of dynamical triangulations. The older one, known as dynamical triangulation (DT), was used to discretize the  functional integral for Euclidean field theory, which on the lattice can be  interpreted as  describing a statistical infinite-dimensional system.
  A more physical and  recent approach is provided by  the Minkowskian dynamical triangulation with triangularized space and discretized global time, called causal dynamical triangulation (CDT) \cite{lukbibl59,lukbibl60,lukbibl61}.
  In the CDT approach one assumes the triangular discretization of curved three-space and the time evolution is discretized in a way consistent with the Hamiltonian   lattice  framework.
  The triangularization in the CDT method is fitted therefore into a space-time layer between two discrete global time values $t$ and $t+\Delta t$, i.e.  similarly  as in nonrelativistic approach we strongly distinguish space and time discretizations.
  It should be added however that in a parallel way to the development of the covariant formulation of LQG,  recently also
  a covariant CDT approach without singling out the global time coordinate has been investigated,   mostly studied in $D=2+1$ QG \cite{lukbibl62}.

  The dynamical lattice approach to QG in the following way addresses  the list of desirable properties of QG models: 
  \begin{description}
  \item{A.\ \ }CDT technique, with variable dynamical triangulation, does not require fixed gravitational three-dimensional background, however the  modeling of discretization in lattice approach does not follow from physical properties of field-theoretic system under consideration.
  Lattice construction is
   a formal regularization method  required also by the efficiency of
   calculational  method.
    In CDT method
     the requirement of background independence should be replaced by the
    independence of obtained results from the choice of imposed discretization scheme. \vspace*{-0.1cm}
  \item{B.\ \ }The diffeomorphism invariance  with the choice of finite lattice size $a$ is approximate.
   Further, it is an important and difficult step to derive   the theory with  consistent  local diffeomorphism invariance   in the limit $a \to 0$.\vspace*{-0.1cm}
  \item{C.\ \ }Because the discrete eigenvalues imitate the `quanta' of quantum space-time, sometimes the discretization in lattice approach to QG is treated as a simulation of physical noncommutativity.
  This argument is however rather misleading, because in lattice approach the choice of discretization is not derived dynamically. \vspace*{-0.1cm}
  \item{D.\ \ }The strong side of the CDT approach is  its good adjustment to computer calculations,  i.e. the method did provide remarkable  calculational results. In particular, following QCD lattice calculations which detected phase transitions in quark-gluon plasma, phase transitions were detected in QG as well.
     Further, in Euclidean lattice calculations (DT approach) one can represent the evolution of dynamical system as a discrete diffusion process, and calculate so-called spectral dimensions at short distances.
  For  $D=4$ QG this spectral dimension  appears to be equal to two, so subsequently one can argue that   one gets the two-dimensional field-theoretic picture of QG at very short distances, what suggests renormalizability.
  \end{description}

  \section{Final remarks}

  We recall again that we did not consider in this note the quite popular description of QG as a low energy effective  sector of quantum (super)string field theory, or
    more recently  as the sector of quantum  $M$-theory.
  The idea of embedding QG in quantum (super)string field theory
   (see e.g.  \cite{lukbibl63,lukbibl64})
    aims at solving at once two basic problems in the theory of fundamental interactions:
   to unify  all interactions and providing finite quantum theory.
   It should be stressed, however, that the basic unsolved questions in the string approach to QG remain: for example  how to incorporate into such an approach the experimentally  well established Standard Model of elementary particles.

  The renormalizability or  finiteness  of a string or $M$-theory  extension of Einstein QG requires the
  use of several post-Einsteinian geometric concepts, namely
  \begin{description}
  \item{i)} additional space-time dimensions ($D=10$ or $D=11$),
  \vspace*{-0.1cm}
  \item{ii)} local supersymmetry,
  \vspace*{-0.1cm}
  \item{iii)} infinite collection of stringy particles (string excitations) which in a somewhat miraculous way leads to the finite interactions of infinite spin multiplets (see e.g. \cite{lukbibl65}).
  \vspace*{-0.1cm}
  \end{description}
 Also strongly linked to the string field theory formulated in AdS space-time  is the very successful idea of  AdS/CDT correspondence \cite{lukbibl66,lukbibl67,lukbibl68}, which provides in a field-theoretic framework the  dual description of gravity and YM gauge theories.
 Such a surprising dynamical picture indicates that at the nonperturbative level QG is holographic, i.e. described in space-time by the degrees of freedom defined on  its boundary \cite{lukbibl68a,lukbibl68b}.
  We recall that a spectacular application of the AdS/CFT
  correspondence is the relation of a weak coupling limit of gravity with a strong coupling sector in YM theory, what was used  to deduce new properties of quark-gluon plasma from known perturbative gravity solutions    (see e.g. \cite{lukbibl69}).

  In Sect.~3 we described three approaches to QG based on different quantization schemes. All these approaches are in continuous progress, but they still do not indicate clearly how a fully satisfactory QG model will appear.
  There is still room for entirely new ideas, for example
  \\ \indent
  {--} Following Penrose one can consider as fundamental the geometry of conformal spinors (twistors), with space-time emerging as  a nonprimary `composite' geometry.
  Twistor ideas have been  studied  already for fifty years (see e.g. \cite{lukbibl70,lukbibl71,lukbibl72}),
    however
   recently an increasing  number of physically important  models  were formulated with the tools of twistor geometry (see e.g. \cite{lukbibl73,lukbibl74,lukbibl75}),
  \\ \indent
  {--} There have been formulated approaches with the pregeometry and emergent gravity concepts.
   As examples of pregeometry can be listed e.g. spin models of Ising type and the emergence of space-time and gravity  due to collective phenomena (see e.g. \cite{lukbibl76}).

  Finally, we stress that what  matters for physicists is the experimental confirmation of QG effects which we expect to be present at very short (Planckian) distances or at a very early stage in the evolution of the Universe.
 The prospects of a detection of such effects favours the  astrophysical measurements.
  Unfortunately, till present time such effects have not been observed, but
   there were proposed many more or less justified
    theoretical predictions of phenomena indirectly related with QG \cite{lukbibl77,lukbibl78} (violation of Lorentz invariance, modification of light velocity, particular anisotropies of CMB, primordial black holes etc.).
  All these proposals are under numerous  studies
   with a common
    hope that the experimental `Planck window' for  QG effects will open in the future.
     The present models of QG are therefore continuously  developing, but
     it is not excluded, that the first confirmation of QG and the construction of a `correct' QG model will come from an unexpected direction (linked with dark matter?).

   In conclusion, one
    can say that the present QG is described by several very promising but still   hypothetical models, which are definitely much more than speculations but less than the long
     awaited
     theory of quantum gravitational phenomena.


    \section*{Acknowledgments}
    {The article is based on talks presented at the 42$^{\rm nd}$ Meeting of Polish Physicists in Pozna\'{n} (September 2013) and at the CERN TH Seminar (November 2013). This research was supported by Polish National Centre of Science (NCN)-Research Project No 2011/01/ST2/03354.}



\end{document}